\documentclass[twocolumn,aps,prb]{revtex4-1}
\usepackage{graphicx}
\usepackage{dcolumn}
\usepackage{bm}
\usepackage{multirow}
\begin{document}
\title{DFT+U study of magnetic order in doped La$_2$CuO$_4$ crystals}

\author{Simon Pesant and Michel C\^ot\'e }

\email{Michel.Cote@umontreal.ca}

\affiliation{D\'epartement de physique et Regroupement qu\'eb\'ecois sur les mat\'eriaux de pointe (RQMP),
Universit\'e de Montr\'eal, C. P. 6128 Succursale Centre-ville,
Montr\'eal (Qu\'ebec) H3C 3J7, Canada}

\begin{abstract}
\small{This article presents the results of several magnetic phases of doped La$_{2-x}$Sr$_x$CuO$_4$ using density-functional theory with an added Hubbard term (DFT+U). Doping factors from $x=0$ to $0.25$ were examined. We found that a bond centered stripe is the magnetic ground state for $x=\frac{1}{8}$ and $x=\frac{1}{4}$. No stable stripe order was found for $x=\frac{1}{6}$. Analysis of the electron density revealed that apical oxygen atoms, those located above and below the copper atoms in the CuO$_2$ planes, hold a non negligible part of the holes at large doping and present a small spin polarization. Finally, the charge reorganization caused by the magnetic stripe modulation was studied for bond centered and atom centered stripes.}
\end{abstract}

\maketitle

\section{Introduction} 

Over the last few years, studies on high-Tc cuprates have reported that the superconducting state competes with magnetic stripe phases.\cite{amtsenecal,compafsc2,Patterson} For low doping, the superexchange~\cite{superexchangepaper} is strong enough to favor an antiferromagnetic order. With increased doping, this interaction is weakened and a superconducting phase can arise. In the La$_{2-x}$Sr$_x$CuO$_4$ system, a sharp decrease in the superconducting temperature is observed for a doping factor of $x=\frac{1}{8}$. At that fraction of strontium substitution, static stripe ordered phases have been detected.\cite{picsstripe} Magnetic phases were also identified for other dopings, up to 0.18.\cite{stripe18,kivelsonrev,stripeminirev}

\begin{figure}[h!]
\centering
\includegraphics[width=0.5 \linewidth]{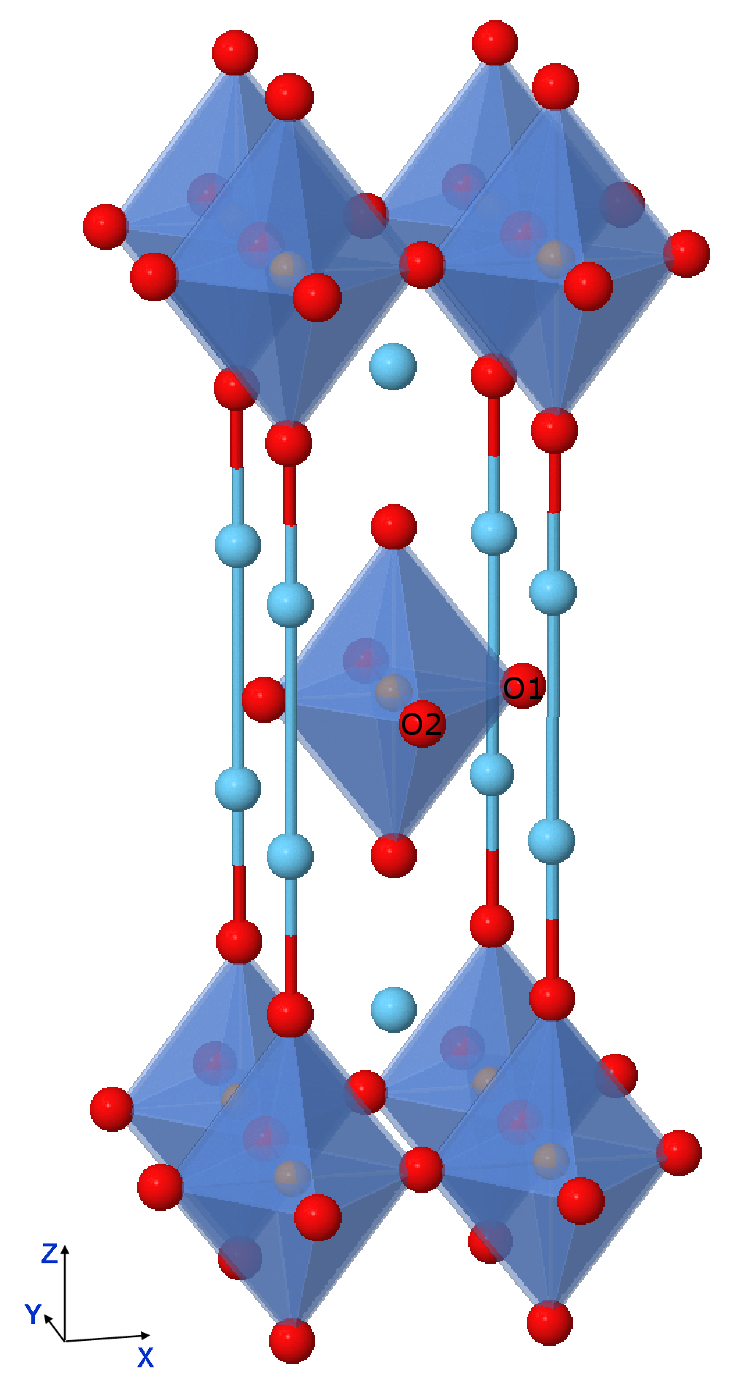}
\caption{\label{lacuostruc}Atomic structure of La$_2$CuO$_4$. Lanthanum atoms are represented by blue spheres, copper atoms by gold spheres and oxygen atoms by red spheres. Apical oxygens are located above and below the copper. In plane oxygens O$_1$ and O$_2$ are in $x$ and $y$ directions in the CuO$_2$ plane.}
\end{figure}

\begin{figure*}[t!]
\centering
\includegraphics[width=0.8 \linewidth]{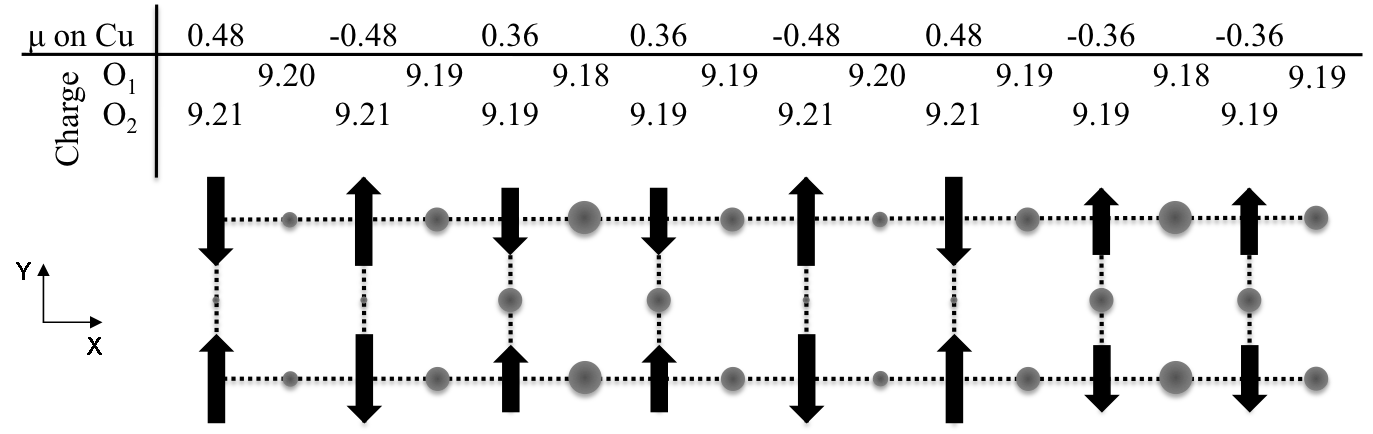}
\caption[ Representation of the bond-centered stripe]{\label{bondcentered} Representation of the bond-centered stripe. The oxygen atoms are represented by circles and the size of each circle is proportional to the hole content on this oxygen atom. For every copper atom, an arrow indicates the orientation of the magnetization of its $3d_{x^2-y^2}$ orbital. On top of the picture are presented respectively the copper magnetizations and the charge on O$_1$ and O$_2$. The values are reported with respect to their position in the stripe. The values were obtained using $U$= 4~eV}
\end{figure*}

\begin{figure*}[t!]
\centering
\includegraphics[width=0.8 \linewidth]{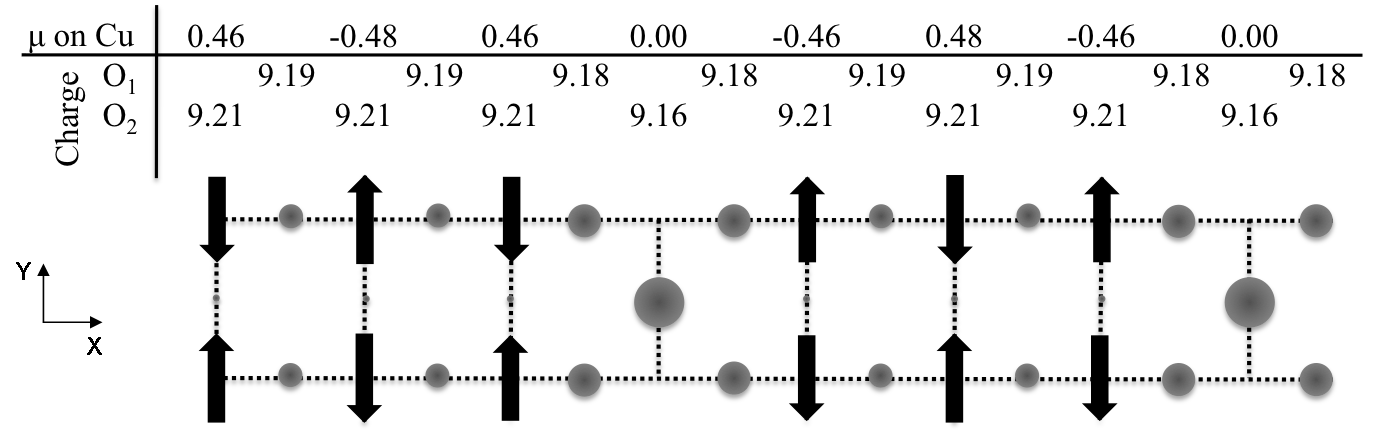}
\caption{\label{atomcentered} Representation of the atom-centered stripe. The oxygen atoms are represented by circles and the size of each circle is proportional to the hole content on this oxygen atom. For every copper atom, an arrow indicates the orientation of the magnetization of its $3d_{x^2-y^2}$ orbital. On top of the picture are presented respectively the copper magnetizations and the charge on O$_1$ and O$_2$. The values are reported with respect to their position in the stripe. The values were obtained using $U$= 4~eV}
\end{figure*}

The atomic structure of the La$_2$CuO$_4$ is depicted in Fig.~\ref{lacuostruc}. The oxygen atoms in the middle of the octahedra and the copper atoms form the copper-oxide plans where stripe ordered phases have been reported. In order to understand the phenomena that yields spin stripes in La$_{2-x}$Sr$_x$CuO$_4$, two commensurate magnetic arrangements were proposed in previous works.\cite{picsstripe,anisimovstripe} They are depicted in Fig.~\ref{bondcentered} and \ref{atomcentered}. Their types of magnetic structure break the translation symmetry of the unit cell. The main difference between these stripes is the location of holes in the structure. In the first one, the holes are located on copper atoms and it is generally referred as an atom-centered stripe (AC). The other configuration is known as a bond-centered stripe (BC) because holes are localized on oxygen atoms between copper atoms in the CuO$_2$ planes.   

It is relatively difficult experimentally to measure the spin configuration in the stripe configurations. To help characterize these phases, we conducted an \emph{ab initio} study on different magnetic structures to determine their relative stability. To this end, we used an extension of density-functional theory that include a Hubbard term within the hamiltonian\cite{Anisimov1997} which we found is necessary to stabilize the stripe configurations. We also looked at these magnetic phases for different doping factors. Precisely, we report the study of the stability of magnetic orders for doping ranging from $x=0$ to 0.25. Bond-centered and atom-centered stripes were considered as well as the antiferromagnetic state. The next section describes the method used and it is followed by a section where we analyze the stability of the different magnetic stripes, the amplitude of the magnetic moments on copper atoms and the atomic charge density for each stripe. Finally, holes distribution is calculated in different stripes and compared to the antiferromagnetic phase with the same doping. It is found that for $x$=0.125 and 0.25, the ground state is the BC stripe and for large doping values, a non negligible part of holes are localized on apical oxygen atoms.

\section{Method\label{secmethod}}

This study has been conducted within the Kohn-Sham approach\cite{kohnsham} of the density-functional theory\cite{hohenbergkohn} (DFT) using the Projected Augmented Waves (PAW)~\cite{pawblochlpaper} formalism as implemented in the ABINIT code.\cite{abinitcodenew,amadonpaw} For the exchange-correlation functional, we used the general gradient approximation of Perdew-Burkes-Ernzerhof.\cite{pbe} To account for the correlation energy of the strongly localized copper $3d$ orbitals, we employed a Hubbard term within the functional (DFT+U).\cite{Anisimov1997} The Hubbard parameters $U$ and $J$ within this method need to be assigned.\cite{note1} We have considered two values for the $U$ term, 4 and 8~eV, with a $J$ term of respectively 0.4 and 0.8~eV. The value of 8~eV for $U$ was chosen because it was determined using a density-functional approach\cite{Anisimov1991} in a previous study of this system\cite{anisimovstripe} using a local density functional for the exchange-correlation energy. Since the present study uses a general gradient approximation for the exchange-correlation functional which is thought to account better for the correlation energy of localized orbitals, we also used the lesser value of 4~eV for the $U$ term. A plane-wave basis set with a maximal kinetic energy of 35 Ha was employed. The Monkhorst-Pack~\cite{monkhorst_pack} k-point grids  for the 8 units supercell, 6 units supercell and 4 units supercell were respectively 4x4x2, 6x6x4 and 8x8x6. Full relaxation of the structures was done in order to obtain their lowest energy configuration. 

\begin{figure}[h!]
\centering
\includegraphics[width=0.8 \linewidth]{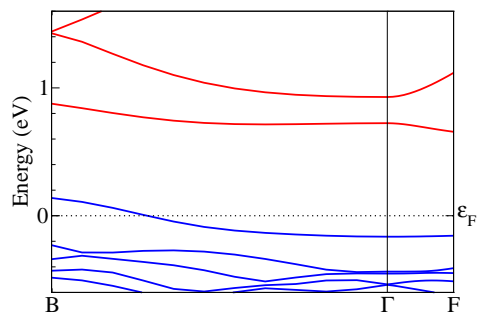}
\caption{\label{bcstripebands} DFT+U band structure of the BC stripe with $U$=4~eV. The band crossing the Fermi level corresponds to the direction perpendicular to the stripe. The other path is oriented along the stripe.}
\end{figure}

The doping of the antiferromagnet La$_2$CuO$_4$ was done by subtracting a number of electrons from the system and adding a neutralizing background. The atomic structures studied were based on the tetragonal cell with space group I4/mmm. This structure gave good results for doped La$_2$CuO$_4$ crystals in a previous work.\cite{giustino} It was also observed that for static stripe ordered phases with neodymium added to pin down the magnetic order, the crystal shows I4/mmm symmetries~\cite{lttphase1, lttphase2} which validate the use of this atomic structure for the study of stripes in La$_2$CuO$_4$. Recently, it was observed that stripes can occur in crystals doped only with barium.~\cite{stripepressure} This indicates that magnetic rare-earth atoms are not required for the appearance  of stripes in La$_2$CuO$_4$ crystals further validating our doping procedure.

\begin{table}[h!]
\centering
\begin{tabular}{|c|c|c|c|c|c|c|}
\cline{2-7}
\multicolumn{1}{c}{} & \multicolumn{2}{|c}{$x = 1/8$}& \multicolumn{2}{|c}{$x = 1/4$} & \multicolumn{2}{|c|}{$x = 1/6$}\\ \cline{2-7}
\multicolumn{1}{c}{} & \multicolumn{1}{|c|}{$U$=4eV} & \multicolumn{1}{c|}{$U$=8eV} &\multicolumn{1}{c|}{$U$=4eV} & \multicolumn{1}{c|}{$U$=8eV} & \multicolumn{1}{c|}{$U$=4eV} & \multicolumn{1}{c|}{$U$=8eV} \\ \hline
BC& 0.065 & 0.282 &0.020 & 0.197 & \multirow{2}{*}{0.009}  & \multirow{2}{*}{0.091} \\ \cline{1-5}
AC  & 0.058 & 0.245 & 0.014 & 0.129 & & \\ \hline
AF  & 0.007  & 0.242 & 0.002 & 0.184 & 0.054 & 0.272 \\ \hline
\end{tabular}
\caption{\label{energytab}Energy differences per copper atom (in eV) between the magnetic phases considered and the normal metallic phase for three doping values. The positive values mean that the magnetic phases are more stable than the normal phase. Results are given for two $U$ values. The antiferromagnetic phase is named AF. Note that for $x=\frac{1}{6}$, only one type of stripe was calculated. See text for details.}
\end{table}

\section{Results and Discussion\label{resultanddiscussion}}

First, we will present the study of the stability of magnetic phases and the copper magnetization. It will be followed by the charge analysis of the stripe phases. Finally, the hole distribution is computed for the different stripes considered. 

\subsection{Total energy stability analysis}

For $x = 0.125$, we constructed a 8 units supercell in which we removed one electron. We first calculated the normal state given by the DFT+U approach without spin polarization. This state is clearly metallic due to the odd number of electrons per primitive cell. We then computed the three magnetic states, the AC and BC stripes and the antiferromagnetic phase, and we calculated the total energy differences between these states and the normal metallic state. These results are reported in Table~\ref{energytab}. We find that the BC stripe has the lowest energy for both values of $U$ used. This indicates that the bond-centered stripe should be the ground state of La$_{1.875}$Sr$_{0.125}$CuO$_4$ which is in agreement with previous results reported in the literature.\cite{anisimovstripe} The band structure of this stripe for $U=4$~eV, shown in Fig.~\ref{bcstripebands}, shows a metallic behavior. The band crossing the Fermi level is in the direction perpendicular to the stripe. Increasing $U$ to 8~eV does not change the metallic property of the crystal but it reduces the bandwidth of the highest occupied band from 0.3 to 0.2~eV. It also increases the band gap between the half filled band and the first completely unoccupied band to 1.9~eV. This is similar to the non-doped antiferromagnetic insulator phase where an increase of $U$ pushes the unoccupied band to higher energy. Furthermore, we note that the atom-centered stripe is the second lowest energy state. Only a small energy difference separates the BC stripe from the AC stripe for both values of $U$ used. We also see that the antiferromagnetic state is barely stable as compared to the normal state for a $U$ value of 4~eV but that the stability of this state is greatly increased for a $U=8$~eV and it is comparable to the AC stripe total energy for this value of $U$. The reason is that an increase of $U$ favors the magnetization of the electrons in the $3d_{x^2-y^2}$ orbital. Finally, we notice that the difference in energy between of the BC stripe and the AC stripe increased when the $U$ term changed from 4 to 8~eV. 

\begin{figure}[h!]
\centering
\includegraphics[width=0.6 \linewidth]{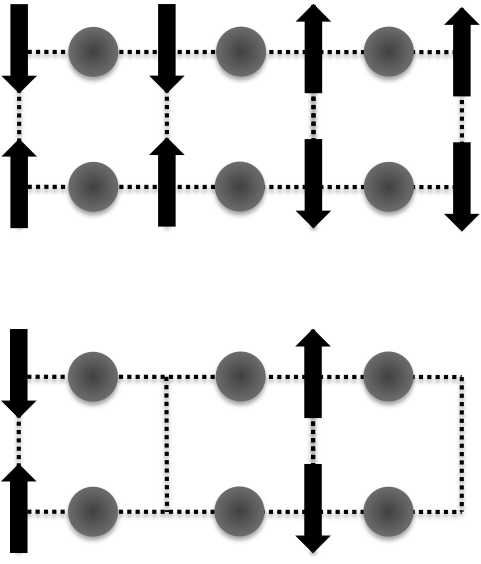}
\caption{\label{stripe4} Representation of the two magnetic stripe orders. The upper panel represents the bond-centered stripe (BC) and the lower panel the atom-centered stripe (AC). The circles represent the oxygen atoms. Arrows, located on the copper sites, indicate the orientation of the magnetization of the $3d_{x^2-y^2}$ orbitals.}
\end{figure}


Using the same approach, we studied other magnetic configurations with different doping levels. First, we constructed a supercell containing 4 primitive cells. Removing an electron from the total density, we obtained a doping fraction per formula unit of $\frac{1}{4}$. We then computed three commensurate magnetic structures, one being the antiferromagnetic state. The other magnetic structures are shown in Fig.~\ref{stripe4}. These magnetic orders are based on the 8 units supercell spin configurations. They represent possible states for La$_{1.75}$Sr$_{0.25}$CuO$_4$. The difference in total energies for these different structures are also reported in Table~\ref{energytab}. Again, the BC arrangement is the lowest energy state. Nonetheless, the energy differences are smaller compared to the case of $x = \frac{1}{8}$. This is due to hole formation in oxygen and copper orbitals that limits the effective spin coupling on copper atoms. 
Nevertheless the DFT+U method predicts that a magnetic order, similar to the BC structure, can be found in La$_2$CuO$_4$ for a doping of $x=0.25$. Interestingly, the AC stripe is not favored compared to the antiferromagnetic state for  this doping and $U$=8~eV. For the smaller value of $U$, the AC phase is more stable than the antiferromagnetic order. This is explained by the important increase in stability of the antiferromagnetic phase for $U$=8~eV compared to $U$ = 4~eV. For this doping factor, the superconducting temperature is almost zero. Consequently, a stripe phase would not generate an important drop in superconducting critical temperature($T_c$). Besides, the results obtained by the DFT+U show that for a $U=4$~eV in copper orbitals, the disparity between the normal metallic and the antiferromagnetic states is 2~meV per copper atom. Because no static antiferromagnetic state is observed for $x=0.25$, the small difference in energy obtained with this value of $U$ seems to describe correctly this system. The BC stripe being an order of magnitude more stable than the antiferromagnetic state, it means that BC stripe could be detected in La$_{1.75}$Sr$_{0.25}$CuO$_4$. It is possible though that this state is fluctuating, like in the case of stripe order in 0.125 doped La$_2$CuO$_4$.  The results with $U$=8~eV show that for a greater Coulomb repulsion, the stripe ordered phase is even more favored. However, for this value of doping, it is reasonable to think that the effective $U$ term would be even smaller than 4~eV which would explain why the metallic state is the most energetically favorable state at that doping.

In order to investigate the stability of magnetic order near the optimal doping ($x\approx$ 0.19), we studied a 6 units long supercell with a doping factor of $\frac{1}{6}$. This time, magnetic orders based on the stripes for $x=0.125$ were not possible without breaking the commensurability of the magnetic unit cell. As a result, we considered a magnetic stripe consisting of three up spins followed by three down spins. We found that this phase was not stable compared to the antiferromagnetic state for the two values of $U$. The AC and BC stripes being incompatible with the commensurability of the supercell, DFT+U predicts that no commensurate stripe order should be detected for $x=\frac{1}{6}$. In this case, the energy difference between the antiferromagnetic state and the normal metallic state is a little higher than for the case $x=0.125$. The difference in doping value being close to 0.125, the antiferromagnetic phase of $x=\frac{1}{6}$ should be similar to 0.125 phase. This will be shown in the next section.

\subsection{Magnetic moments}

For each doping value, the main difference between the stripe configurations is the region where holes are located. Depending on where the holes are based in the CuO$_2$ plane, it modifies the interactions between copper and oxygen atoms. As mentioned before, it affects the superexchange interaction resulting in a modulation of the magnetic moments throughout the supercell. To analyze the magnitude of the moments, we can split the stripes in antiferromagnetic and ferromagnetic parts. The BC stripe can be described by alternating antiferromagnetic and ferromagnetic regions. On the other hand, the AC stripe contains two antiferromagnetic domains separated by non-polarized walls.  We found that, for $x=0.125$, see Fig.~\ref{bondcentered} and \ref{atomcentered}, the antiferromagnetic parts for the BC as well as for the AC stripes have magnetic moments very close to the antiferromagnetic phase with the same doping factor. In fact, the copper magnetization in the antiferromagnetic phase are 0.45 and 0.63 $\mu_B$ for $U$ = 4~eV and 8~eV respectively. At zero doping, the experimental value is 0.6~$\mu_B$ which is well reproduced with a $U$ of 8~eV, however, for this value of the Coulomb repulsion, the antiferromagnetic doping region is too large. The variation in magnetic moments for the BC stripe is 0.12~$\mu_B$ for $U=4$~eV (0.05~$\mu_B$ for $U=8$~eV). Nonetheless, each copper atom has a non-zero magnetization. For the AC stripe, this variation is greater because of the two non-polarized atoms. But the difference of absolute values for magnetic moments between polarized atoms in the AC stripe is very low. Finally, we see that the DFT+U approach can give rise to different modulations of spins in La$_2$CuO$_4$ doped crystal. For $x=0.25$, the antiferromagnetic phase has a magnitude of spin polarization on the copper atom of 0.31~$\mu_B$ for a Coulomb repulsion of 4~eV. This value is lower than those for $x=0.125$. For $x=\frac{1}{6}$, we found a magnetization on the copper atom of 0.47~$\mu_B$ which is as high as for $x=0.125$. This is consistent with the high stability of the antiferromagnetic phase of the 6 units supercell.

\subsection{Charge analysis and hole distribution}

\begin{figure}[h!]
\centering
\includegraphics[width=0.8 \linewidth]{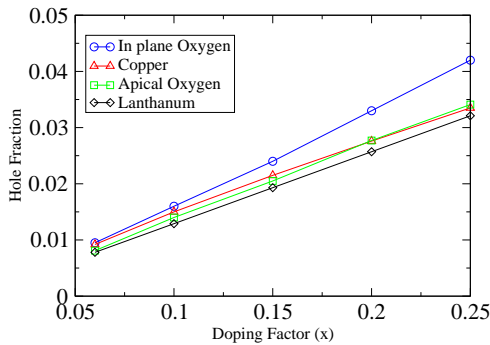}
\caption{\label{graphcharge} Location of holes in the antiferromagnetic phase of doped LaCuO for $U$=4~eV. Lines are guides to the eyes.}
\end{figure}

The spin configurations in the supercells can be understood by the hole distribution, \emph{i.e.} the localization of holes in the CuO$_2$ planes. In order to analyze this distribution through the crystal, we calculated the Bader charges\cite{baderlacuo} in these systems with a $U$ of 4~eV. This method splits up the electronic density associated to a specific atom. Bader volumes, containing atomic domains, are defined by zero flux surfaces of the electronic density. First, we computed the antiferromagnetic phases for $0\leq x \leq 0.25$. As mentioned before, with a value of $U$=4~eV, the DFT+U can find an antiferromagnetic phase even for high doping ($x=0.25$). For lower $U$, the predicted antiferromagnetic doping region can be reduced, however, the resulting moments for the non-doped system will be much lower than the experimentally observed value. It seems that within the DFT+U approach, the $U$ value should be lowered as doping increases to account for the experimental phases observed. In the present study, we considered a $U$ term of 4~eV for all doping, which seems a reasonable value. To obtain the location of holes in the crystal, we take the difference between the non-doped system and the doped one. Fig.~\ref{graphcharge} shows the increase of hole content on copper, apical oxygen, in-plane oxygen and lanthanum atoms with respect to the doping factor $x$. When adding up the hole fraction of each atom in the primitive cell at a given doping factor, we note that an important part of the charge, about 50\% of the total hole content, is localized on the apical oxygen and lanthanum atoms. Consequently, atoms outside the CuO$_2$ are also affected by doping. It was found in a previous work~\cite{dopingbonn} on YBaCuO crystals that the c-axis length is related to $T_c$ and hence to the doping level. The major cause of c-axis variation with respect to hole doping being the apical oxygen positions, our results confirm the importance of atoms outside the plane in holes distribution and, indirectly, to superconductivity. The charge on lanthanum is justified by the fact that even for real substitution of lanthanum by strontium, a little portion of the hole introduced into the structure should stay on the dopant. Moreover, we can see that the fraction of holes centered around the copper atoms is of the same magnitude or lower than the others atoms. In fact, the Bader analysis reveals that the majority of the holes will be injected in oxygen orbitals. This is consistent with experiments \cite{tranquada1987} and the electronic properties of charge transfer insulators where the oxygen orbitals lie just below the Fermi level. Finally, we note that hole formation increases more rapidly on in-plane oxygen than on other atoms. This means that for large values of doping, hole content should increase faster in the copper oxide planes then in the rest of the system. The hole formation in CuO$_2$ planes also explains the decrease of magnetic moments on the copper atoms with increased doping.

We also conducted a Bader analysis on spin densities for the same doping as the previous charge analysis. From this, we found that apical oxygen atoms are slightly spin polarized. In fact, the non-doped crystal shows a magnetic moment of 0.01~$\mu_B$ on the apical oxygen atoms. This small spin polarization is also found in the $3d_{z^2}$ orbitals of the copper atoms. This value doubles for $x=0.25$ system. The increase follows the hole content in these orbitals. This phenomenon, in opposition to the antiferromagnetism in the CuO$_2$ plane, is due to the direct exchange between $2p_z$ orbitals of apical oxygen atoms and copper $3d_{z^2}$ orbitals. A Wannier functions decomposition of the band structure for these systems reveals that $2p_z$ and $3d_{z^2}$ orbitals present the second largest overlap. The first one being the overlap of the $3d_{x^2+y^2}$ and $2p_x$ of the in-plane oxygen atoms. In addition, we found that this polarization is noticeable only for values of $U$ higher than 2~eV. 


For the stripe magnetic orders, the Bader analysis shows the break of translation symmetry. For the BC and the AC stripes, the charge modulation shows a commensurability of 4 unit cells as observed in experiments.\cite{picsstripe} For both types of stripes, we found that the charge modulation is practically all on the oxygen atoms. The values for the charges obtained for the in-plane oxygen atoms are reported in Fig.~\ref{bondcentered} and Fig.~\ref{atomcentered}. Considering the BC stripe first, we found that it does not show a noticeable modulation of copper charges. The atoms in the antiferromagnetic parts have a charge of 27.94 electrons and copper atoms in the ferromagnetic domains have a smaller value by only 0.01 electron. Contrary to the antiferromagnetic phase, stripes present different atomic charges for both in-plane oxygen atoms. The CuO$_2$ planes now have two different types of oxygen atoms. One is located between the copper atoms along the 8 units supercell stripe (O$_1$) and the other is parallel to the magnetic configuration (O$_2$) (see Fig.~\ref{lacuostruc}). This is another evidence of the symmetry breaking induced by stripe ordered phases. O$_1$ atoms have their maximum charges between copper atoms in the middle of the antiferromagnetic regions. On the contrary, their minimal charge, which is at the same time their maximum in hole content, is centered on the oxygen atoms between copper atoms in the ferromagnetic wall. Consequently, holes are favored in the ferromagnetic domains. For O$_2$ atoms, the charge maximum and minimum follow the trend of the magnetization of the copper atoms, meaning that the larger charge values are found in antiferromagnetic chains with largest value of magnetization, which is essentially the same values as for the non-doped system, and the lowest charge values are found in the lower magnetization chains. Also, the total charge of O$_2$, compared to O$_1$, is greater by 0.08 electron over the 8 units cell. It indicates that BC stripes favor hole formation on oxygen atoms along the magnetic modulation. With these results, we see that the antiferromagnetic region is less doped compared to the ferromagnetic part. A charge modulation on apical oxygen atoms is also obtained. It is due to modulation of magnetization on copper atoms and also to a small tilt of the octahedra of the CuO$_2$ planes.


The same analysis can be done for the AC stripe. First, the copper charges are almost the same for every atoms in the antiferromagnetic regions. For this stripe, all copper atoms have the same charge value of 27.94 electrons. This time, the modulation of copper magnetization is not followed by a charge variation. Nonetheless, the magnetic ordering still splits the charge symmetry for O$_1$ and O$_2$. For O$_1$ and O$_2$, the modulation of charges is only substantial for oxygens near the non magnetic walls. Particularly, the minimum of O$_2$ charge is on the atom bonded to the non magnetic copper and the minimal charges of O$_1$ are found on each side of the non magnetic walls. The maximal charge variations are 0.01 and 0.05 electron for the first and second type of in-plane oxygen. In addition, the difference of the total charge between O$_1$ and O$_2$ is 0.1 electrons over the whole stripe which is slightly larger than for the BC stripe. For the AC stripe, the Bader analysis shows that oxygen atoms around non magnetic copper can be compared to those of highly doped systems, especially for O$_2$. Finally, the difference between BC and AC stripes, in the CuO$_2$ plane, is the range where holes are located. In fact, for AC stripes, an appreciable part is localized on oxygen atoms in the non magnetic chains whereas for BC stripes, the charge is more evenly distributed. 



If we now consider the bond length between copper and O$_1$ atoms ($x$ axis), we see that it varies depending on the charges on the neighboring oxygen atoms. For the BC stripe, the distance between copper and oxygen atoms along the $x$ direction ranges between 1.89 and 1.91~{\AA}. The maximal bond length is found for copper atoms linked to an O$_1$ atom that has a charge of 9.19 electrons and to an O$_2$ atom with a charge of 9.21 electrons. The charge disparity between the two types of oxygen atoms surrounding the same copper atom modifies the bond length. The longer bond length of the O$_1$ atom indicates a weaker bonding to the copper atom compared the O$_2$ atom. Moreover, the other O$_1$ atoms in the antiferromagnetic regions show bond lengths of 1.91~{\AA} with their nearest copper atoms, the difference in charge between O$_1$ and O$_2$ atoms being insufficient to noticeably modify the bond lengths. On the contrary, the 1.89~{\AA} bond length found in the ferromagnetic domains are associated to O$_1$ atoms having the lowest atomic charge. The reason is that, due to the symmetry in the $y$ axis, Cu-O$_2$ bond length are fixed to 1.9~{\AA} as in the antiferromagnetic zone. The O$_2$ atom having a smaller atomic charge in ferromagnetic regions than in antiferromagnetic regions, the fixed bond length of O$_2$ atoms prevents this atom from being closer to the copper atom. As a result, it favors the reduction of the bond length between copper atoms and its closest O$_1$ atom. Finally, the BC stripe phase affects simultaneously the copper magnetic moments, the charge distribution and the CuO$_2$ plane structure to lower the total energy of the system. 
In the same way, bond lengths in the AC stripe are modulated along the $x$ direction. For this phase, the maximum bond length is still 1.91~{\AA} and is found in the non-magnetic wall. As discussed before, the larger bond length obtained for Cu-O$_1$ link is explained by difference between O$_1$ and O$_2$ atomic charges, when the O$_2$ atom has the maximal atomic charge (9.21 electrons). In addition, the smallest bond length is also found in the non-magnetic wall. In fact, the AC stripe presents the lowest atomic charge for in plane oxygen atoms (9.16 electrons) when compared to the BC stripe. This important concentration of hole localized on an O$_2$ atom, combined to the fixed Cu-O$_2$ bond length (1.9~{\AA}) result in a bond length of 1.88~{\AA}, similarly to the BC stripe. The smallest bond length found in the AC stripe, compared to its analogue in the BC stripe, is accounted by the higher hole content in the O$_2$ atoms of the AC stripe. 
The fact that BC stripe is more stable than the AC stripe indicates that the structure favors shorter bond lengths modulation along the $x$ axis. The proximity of the O$_1$ atom and the copper atom in the 1.88~{\AA} bond of BC stripe come with a small increase in the total energy of the system due to Coulomb repulsion, which explains, partially, the higher energy of this phase compared to the BC stripe.

The Bader analysis on the BC structure can be compared to the antiferromagnetic state of the non-doped system to give a more precise location of holes in this stripe. The study of Bader charges in antiferromagnetic phases showed that copper atoms only hold a fraction of the total holes injected in the structure. This is still the case for BC stripes. In fact, copper atoms possess a slightly larger hole fraction, 0.18 hole, compared to the antiferromagnetic phase with the same doping. It can be accounted by the variation of the copper magnetic moments, which tend to result in a increased hole concentration on low magnetization atoms. In the case of in-plane oxygen atoms, they hold a major part of the total hole content, 0.37 hole, and it is distributed almost evenly with a larger hole content on O$_1$ oxygen atoms between the two copper atoms in the ferromagnetic region. In this type of stripe, holes are more localized on O$_1$ oxygen atoms as compared to O$_2$ oxygen atoms by about 33\%. As mentioned before, the modulation of magnetization induces a reorganization of holes in the system. The charge analysis revealed that the BC stripe creates a charge modulation for both types of in-plane oxygen. Moreover, this charge modulation creates hole rich and hole poor regions in the CuO$_2$ plane.        




The AC stripe has a smaller amount of hole on their copper atoms, compared to BC configuration, but a larger portion of hole is located on oxygen atoms in the CuO$_2$ planes. The portion of hole doping on copper atoms is reduced to 0.14, comparable to the antiferromagnetic phase. On the contrary, the in-plane oxygen atoms share 0.43 hole, which is higher than in the BC and antiferromagnetic phases. The high value of hole content in the non magnetic wall and oxygen atoms nearby creates a large concentration of hole in this region. In fact, an amount of 0.15 hole is found on these oxygen atoms. This represents a quarter of the total hole contained in the CuO$_2$ plane. The localization of hole is more important in this stripe, where non magnetic walls hold a large fraction of hole. 

Finally, for x=0.125, the BC stripe order is found to be the ground state of the system. From this and the charge analysis, we can conclude that this doped phase of La$_2$CuO$_4$ is more stable than for configurations inducing hole localization in small regions of the CuO$_2$ plane. 

\section{Conclusion}

To conclude, several configurations of La$_2$CuO$_4$ were studied, for hole doping ranging from 0 to 0.25, using DFT+U. In the cases where supercells allow a charge order half the length of the magnetic order, a bond centered stripe is found to be the ground state of the system. The cases $x=\frac{1}{4}$ and $x=\frac{1}{8}$ satisfy this condition. For a doping of $\frac{1}{6}$, no stripes are energetically favored and the lowest energy system is the antiferromagnetic state. This configuration does not allow the same pattern as $x=\frac{1}{4}$ or $x=\frac{1}{8}$. Moreover, a charge investigation on doped phases of antiferromagnetic La$_2$CuO$_4$ indicates that holes are located mainly on oxygen atoms. Also, study of the spin density found that apical oxygen atoms have a small magnetization, caused by the overlap of $2p_z$ and $3d_{z^2}$. Furthermore, hole formation is found to be greater for in-plane oxygens, compared to copper atoms. For stripes, the modulation of magnetic moments on the copper atoms can be understood by the charge reorganization of the oxygen atoms of the CuO$_2$ planes. This means that stripe ordered phases deform lightly the crystal to favor a different magnetic order than the usual antiferromagnetic conformation. Also, it was shown that stripe orders favor the vicinity of hole poor and hole rich regions. This reorganization of the electronic density should induce modification to the phonon spectra for doping factor allowing stripe phases.
 
\begin{acknowledgments}

We are grateful for helpful discussions with Andr\'e-Marie Tremblay and Andrea Bianchi. This work was supported by grants from the FQRNT and the CRSNG. The calculations were done using computational resources provided by the R\'eseau qu\'eb\'ecois de calcul de haute performance (RQCHP). We are grateful to Yann Pouillon for helpful and valuable technical support with the build system of ABINIT. 
\end{acknowledgments}


\newcommand{\SortNoop}[1]{}

\end{document}